\documentclass[twocolumn]{aastex63}
\usepackage[normalem]{ulem}   
\usepackage{nameref}
\usepackage{hyperref}
\usepackage{amsmath}
\usepackage{amssymb}
\usepackage{mathrsfs}
\usepackage{animate}
\usepackage{graphicx}
\usepackage{rotating}
\usepackage{bbm} 
\usepackage{amsthm}        
\usepackage{comment}

\usepackage{svg}
\usepackage[outline]{contour}

\usepackage{subfigure}
\usepackage{tikz}
\usepackage{mdframed}

\usepackage{lineno}

\definecolor{forestgreen}{rgb}{0.13, 0.55, 0.13} %forestgreen5web)
\newcommand{\p}[1]{{\color{magenta}{#1}}}

\newcommand{\cs}{c_\mathrm{s}}

\newcommand{\game}{\gamma_\mathrm{e}}
\newcommand{\gamp}{\gamma_\mathrm{p}}
\newcommand{\gams}{\gamma_\mathrm{s}}
\newcommand{\gamso}{\gamma_\mathrm{s0}}
\newcommand{\gamsl}{\gamma_\mathrm{s1}}
\newcommand{\gampo}{\gamma_\mathrm{p0}}
\newcommand{\gampl}{\gamma_\mathrm{p1}}
\newcommand{\gameo}{\gamma_\mathrm{e0}}
\newcommand{\gamel}{\gamma_\mathrm{e1}}

\renewcommand{\ne}{n_\mathrm{e}}

\newcommand{\Ps}{P_\mathrm{s}}

\newcommand{\rc}{r_\mathrm{c}}
\newcommand{\Te}{T_\mathrm{e}}
\newcommand{\Teo}{T_\mathrm{e0}}
\newcommand{\Tp}{T_\mathrm{p}}
\newcommand{\Tpo}{T_\mathrm{p0}}
\newcommand{\Ts}{T_\mathrm{s}}
\newcommand{\Tso}{T_\mathrm{s0}} 
\newcommand{\Tsc}{T_\mathrm{sc}}
\newcommand{\Tpc}{T_\mathrm{pc}}
\newcommand{\Tec}{T_\mathrm{ec}}
\newcommand{\Tpmax}{T_\mathrm{p,max}}

\newcommand{\xs}{x_\mathrm{s}}

\newcommand{\riso}{r_\mathrm{iso}}
\newcommand{\risoe}{r_\mathrm{iso|e}}
\newcommand{\risop}{r_\mathrm{iso|p}}
\newcommand{\risos}{r_\mathrm{iso|s}}
\newcommand{\rpolys}{r_\mathrm{poly|s}}
\newcommand{\rpolyp}{r_\mathrm{poly|p}}
\newcommand{\rpolye}{r_\mathrm{poly|e}}

\newcommand{\ntil}{\Tilde{n}}

\newcommand{\ntils}{\Tilde{n}_\mathrm{s}} 
\newcommand{\sums}{\sum_\mathrm{s = \{p,e\}}}

\newcommand{\Br}{B_\mathrm{r}}

\newcommand{\rs}{r_\mathrm{\odot}}
\newcommand{\rss}{r_\mathrm{ss}}

\newcommand{\fss}{f_\mathrm{ss}}

\renewcommand{\ne}{n_\mathrm{e}}

\newcommand{\ratioC}{\mathrm{C}^{6+} / \mathrm{C}^{5+}}
\newcommand{\ratioO}{\mathrm{O}^{7+} / \mathrm{O}^{6+}}

\contourlength{0.3pt}
\contournumber{10}

\newcommand{\M}{\mathrm{M}}

\renewcommand{\ne}{n_\mathrm{e}}

\newcommand{\TeO}{T_\mathrm{e|O}}
\newcommand{\TeC}{T_\mathrm{e|C}}

\newcommand{\rf}{r_\mathrm{f}}

\shorttitle{Generalized Two Thermal Regimes Approach : Bipoly Fluid Modeling}
\shortauthors{Dakeyo et al.}
\begin{document}

\title{Generalized Two Thermal Regimes Approach : Bipoly Fluid Modeling}
%\title{Radial Iso-poly Fluid Models of Solar Winds Constrained by Parker Solar Probe and Helios Measurements}

\correspondingauthor{Jean-Baptiste Dakeyo}
\email{jbdakeyo@hotmail.fr}

\author[0000-0002-1628-0276]{Jean-Baptiste Dakeyo}
\affiliation{Space Sciences Laboratory, University of California, Berkeley, CA, USA}

\author[0000-0001-8215-6532]{Pascal D\'emoulin}
\affiliation{LIRA, Observatoire de Paris, Universit\'e PSL, CNRS, Sorbonne Universit\'e, Universit\'e de Paris, 5 place Jules Janssen, 92195 Meudon, France}
\affiliation{Laboratoire Cogitamus, 75005 Paris, France}

\author[0000-0003-4039-5767]{Alexis Rouillard}
\affiliation{IRAP, Observatoire Midi-Pyrénées, Universit\'e Toulouse III - Paul Sabatier, CNRS, 9 Avenue du Colonel Roche, 31400 Toulouse, France}

\author[0000-0001-6172-5062]{Milan Maksimovic}
\affiliation{LIRA, Observatoire de Paris, Universit\'e PSL, CNRS, Sorbonne Universit\'e, Universit\'e de Paris, 5 place Jules Janssen, 92195 Meudon, France}

\author{Alice Chapiron}
\affiliation{IRAP, Observatoire Midi-Pyrénées, Universit\'e Toulouse III - Paul Sabatier, CNRS, 9 Avenue du Colonel Roche, 31400 Toulouse, France}

\author[0000-0002-1989-3596]{Stuart Bale}
\affiliation{Space Sciences Laboratory, University of California, Berkeley, CA, USA}
\affiliation{Physics Department, University of California, Berkeley, CA, USA}

\begin{abstract}
The isopoly bi-fluid approach assumes an isothermal evolution of the solar wind near the Sun up to the radial distance $\riso$, followed by a polytropic evolution constrained by the observed polytropic indices.  This approach provides a more accurate model of the interplanetary properties of the solar wind ($u$, $n$, $\Tp$, $\Te$) and their radial evolution \citep{dakeyo2022,dakeyo2024b}. In this article, we present an improvement of the isopoly approach by considering a generalized two thermal regime approach, embedding two distinct polytropic evolutions, the "bipoly" modeling. To demonstrate the capability of the approach, the models are fitted to both interplanetary and coronal observations, all classified by wind speed population in the spirit of \citet{maksimovic2020}. The set of observations used as constraints are coronal temperatures inferred from charge-state ratio observations from Solar Orbiter, and interplanetary measurements from Helios and Parker Solar Probe. The relaxation of the isothermal criteria in the near-Sun region permits to significantly improve the fast wind acceleration for low coronal temperature conditions.  In summary, the new model matches closely the observational constraints: the coronal temperature and the radial evolution of the wind properties ($u$, $n$, $\Tp$, $\Te$) in the interplanetary medium, and this for all the wind speed populations.
\end{abstract}


%%%%%%%%%%%%%%%%%%%%%%%%%%%%%%%
%%%%%%%%%%%%%%%%%%%%%%%%%%%%%%%
\section{Introduction}
%%%%%%%%%%%%%%%%%%%%%%%%%%%%%%%
%%%%%%%%%%%%%%%%%%%%%%%%%%%%%%%
% Effect of f(r) on heating
The solar wind model of \cite{parker1958} has been widely used to interpret the in-situ and source properties of the solar wind. Parker's seminal work assumed an isothermal corona expanding in the form of a hydrodynamic flow escaping solar gravity. This isothermal assumption is known to break down beyond the solar corona as the bulk flow expands into the interplanetary medium. Space missions such as Helios have found that the temperature of the solar wind decreases away from the Sun \citep{Schwartz1983radial}. 

For an adiabatic spherical expansion with $\gamma = 5/3$, the expected temperature decrease is $r^{-4/3}$. The measured temperature gradient however exhibits a decrease closer to $r^{-1}$ for protons, and of $r^{-0.5}$ for electrons \citep{maksimovic2020, dakeyo2022, halekas2022}. The solar wind is heated in the interplanetary medium, and in a specie-dependent manner. 

In our previous work \citep{dakeyo2022} we proposed the combination of an isothermal and polytropic model solution of the hydrodynamic Parker solar wind. This approach supposed an isothermal evolution of the plasma in the solar corona, combined with a polytropic expansion at greater heliocentric distances, well beyond the sonic point. The model was constrained by interplanetary measurements and classified by wind speed populations in the spirit of \cite{maksimovic2020}. It could reproduce the macroscopic evolution of wind properties, such as the bulk speed $u$, the proton $\Tp$ and electron $\Te$ temperatures and the density $n$, in the interplanetary medium. In the coronal part the temperatures necessary to obtain realistic interplanetary fast wind solutions were in excess of  5.5 MK which is well above the observed $\sim$ 1 - 2 MK derived from coronal spectroscopy \citep{Cranmer1999spectroscopic}.
 
An alternative estimate of coronal temperatures can be obtained from in-situ measurements of heavy ions by considering their freeze-in temperatures \cite{ko1997}. The ionization level of heavy ions is set in the low corona primarily through collisions with hot electrons. This makes the density ratio between different levels of ionic ionization to be directly related to electronic temperatures in the collisional region of the low atmosphere. Studies based on the analysis ion charge-state ratios measured in the solar wind show that the source electron temperature is globally anticorrelated with the wind speed. The source temperatures are in the range of $ \Te \sim$ 1.5 - 2 MK for the slow wind and $\Te \sim$ 1 MK for the fast wind \citep{landi2012, Xu2015, Wang2016}. This is at odds with the isothermal Parker wind model since low coronal temperatures should reduce the thermal pressure gradient responsible for the acceleration of the solar wind.

%Regarding the solar wind evolution in the near Sun region,
Remote observations of the solar corona combined with modeling of its magnetic field show that the magnetic pressure balance defines the expansion rate of the magnetic flux tubes \citep{kopp_holzer1976}. Since the magnetic pressure of the closed magnetic field (traced by the coronal loops) disappears with height, transverse magnetic pressure gradient force a significant expansion of the open magnetic field in the lateral direction. The escaping plasma experiences a super radial expansion in the solar corona, i.e. greater than the spherical expansion. This effect can be included into the solar wind modeling by considering the expansion factor $f(r)$, which describes the expansion rate of a given flux tube. \\

The coronal magnetic field is frequently computed with a potential field extrapolation of the measured photospheric field, assuming that the field becomes open and radial at a spherical source surface of typical radius $r=\rss = 2.5\,\rs$). \cite{wang_sheeley1990} noticed an anticorrelation between the solar wind speed $v$ measured near 1 au and $f(\rss)$ for observations describing large spatial scales taken over several solar cycles. Several complementary studies have looked into this anticorrelation, and have shown that the anticorrelation is also sensitive to the nature of the source, and parameters such as the distance of the derived photospheric point of magnetic connectivity  from the center of the source coronal hole \cite[e.g.][]{arge2000,riley2015}. 

%Another recent complementary study by 
\cite{dakeyo2024b} revisited this $v$ -- $f(\rss)$ relationship based on magnetic connectivity study of wind measurements by Solar Orbiter, and their isopoly fluid model. They highlighted that the $v$ -- $f(\rss)$ anti-correlation is mainly present for the wind component originating in polar coronal holes, i.e. high latitude sources and for widely extended coronal hole structures. Moreover \cite{dakeyo2024b} relate a fraction of the fast solar wind measured in the ecliptic plane to coronal sources at low latitudes associated with large expansion factors. This implies that the expansion factor does not intrinsically set the terminal wind speed. Based on \cite{kopp_holzer1976} work, \cite{dakeyo2024b} proposed that the flow regime in the low corona strongly regulate part of the wind acceleration rate. More specifically, the flow regime could be separated in two classes that they called "f-subsonic" and "f-supersonic" solutions of the solar wind, respectively fully subsonic and partially supersonic in the super expansion region. The f-supersonic solutions are associated with significant acceleration due to the so-called deLaval effect. The super radial expansion of flux tubes provides an extra acceleration to reach faster wind speed. However, these computed isopoly models did not fully agree with source temperatures given by coronal spectroscopy or ion charge-state ratios \citep{ko1997, Xu2015, Wang2016}.

In this paper, we investigate to what extent the isopoly model can account for observed source temperatures, while still reproducing the properties of wind populations in the interplanetary medium considered in \cite{dakeyo2022} and for the expansion factors of\cite{dakeyo2024b}. 
For this purpose we determine from the Solar Orbiter observations, the electronic coronal temperatures in Section \ref{subsec:state_charge_ratio_convert_Teo}, and use them in Section \ref{sec:isopoly_state_charge} to constrain the isopoly equations of \citet{dakeyo2024b} in the near Sun region. We show that the deduced isopoly models are not in agreement with both detailed coronal and interplanetary plasma observations. This leads us in Section \ref{sec:two_thermal_regime_model}, to propose a generalized two thermal regime approach instead of the isopoly modeling by relaxing the isothermal assumption in the solar corona.

%%%%%%%%%%%%%%%%%%%%%%%%%%%%%%%%%%
%%%%%%%%%%%%%%%%%%%%%%%%%%%%%%%%%%
\section{Interplanetary observations and coronal proxy observations}
\label{sec:observations_descript}
%%%%%%%%%%%%%%%%%%%%%%%%%%%%%%%%%%
%%%%%%%%%%%%%%%%%%%%%%%%%%%%%%%%%%

%%%%%%%%%%%%%%%%%%%%%%%%%%%%%%%%%%
%%%%%%%%%%%%%%%%%%%%%%%%%%%%%%%%%%
\subsection{Wind In-situ Measurements}
\label{subsec:wind_bulk_obs}
%%%%%%%%%%%%%%%%%%%%%%%%%%%%%%%%%%
%%%%%%%%%%%%%%%%%%%%%%%%%%%%%%%%%%

The interplanetary observations used to constrain the modeling of the bulk velocity $u$, proton temperature $\Tp$, electron temperature $\Te$, and density $\ne$ are the same as in \cite{dakeyo2022}. The solar wind data was classified into five wind speed populations based on the terminal wind speed. The median wind speed profiles include solar minimum period measurements from Helios 1 and Helios 2 from 1974 to 1977 \citep{1981_ref_helios}, and from Parker Solar Probe for encounters E4 to E9 \citep{fox2016}. See \cite{dakeyo2022} for more details. We use this data set to constrain the interplanetary radial evolution of our models.

The charge-state ratio measurements used to constrain coronal solar wind modeling are obtained from in-situ measurements taken by the Heavy Ion Sensor (HIS) instrument onboard Solar Orbiter \citep{owen_SWA2020}. The solar wind bulk velocity is associated with each HIS time observation using measurements from the Proton Alpha Sensor (PAS) instruments. Since HIS has a lower sampling rate than PAS, the corresponding PAS velocity is calculated as the average velocity over the interval of HIS time sampling ($\approx$ 10 minutes).

%%%%%%%%%%%%%%%%%%%%%%%%%%%%%%%%%%
%%%%%%%%%%%%%%%%%%%%%%%%%%%%%%%%%%
\subsection{Charge-State ratio as Coronal Constraints}
\label{subsec:state_charge_ratio_convert_Teo}
%%%%%%%%%%%%%%%%%%%%%%%%%%%%%%%%%%
%%%%%%%%%%%%%%%%%%%%%%%%%%%%%%%%%%

% Limit of spectroscopy obs to estimate T0
Remote-sensing observations, such as coronal spectroscopy can provide an estimate of the plasma local temperatures and densities. However, only a subset of solar wind species are observable by spectroscopy, such as neutral hydrogen that charge-exchanges with coronal protons and specific lines of heavy ions such as Oxygen VI \citep{Cranmer2002coronal}. \\

% charge-state ratio introduction %   as proxy of Teo
Based on \cite{ko1997} work, the densities of ions measured in-situ can be used to determine the source electron temperatures, i.e. in regions where the coronal plasma is not yet collisionless. This is based on the property that in the collisional corona, the heavy ions become increasingly ionized by electrons as ambient temperature and therefore electron kinetic energy rises.
The hotter, the denser and the more time the electrons remain in the collisional corona, the higher the ionization level of the atoms. 
Using the known ionization and recombination rates of different ion species, it is possible to determine $\Te$ in the corona as a function of the ratio of ionic fractions. % for a given atomic specie.
This way $\Te$ can be determined in the corona below a 'freezing height' located at the radius $\rf$, where collisional frequencies are too low to change ionization levels. The ionic fraction $y_i$ of the ion with charge $+i$ is expressed by~:
  \begin{align}
  y_i \equiv \, \frac{n_i}{\sum_{i=0}^{Z} n_i }.
  \label{eq:Ko1997_y_ions_def}
  \end{align}
where $Z$ is the number of electrons for the neutral atom.
When ions are in ionization equilibrium, such as in the hydrostatic low corona, the ionic fraction depends exclusively on the ionization and recombination rates, $C_i$ and $R_i$ respectively (from charge $i$ to $i+1$ and $i+1$ to $i$, respectively)~:
\begin{align}
    \frac{y_{i+1}}{y_i} = \frac{C_i}{R_i}
    \label{eq:condition_ionic_fraction_equilibrium}
\end{align}
where both $C_i$ and $R_i$ are function of $\Te$ (and other parameters such as the electron density). Higher up in the corona, the drop in plasma density in conjunction with the higher radial speed of the accelerating solar wind typically force the ionization time to exceed the expansion time, and the charge state ratio does not change significantly. 
The heliocentric radial distance at which the ratio remains constant is called the "freeze-in" radius. Therefore, in-situ measurements of the charge-state ratio provide an estimate of $\Te$ at the freeze-in radius. For more details, please refer to \cite{ko1997}.

% Estimate of relation charge-state ratio Teo
A Python code available at \hyperref[https://fiasco.readthedocs.io/en/stable/api/fiasco.Ion.html]{\url{https://fiasco.readthedocs.io/en/stable/api/fiasco.Ion.html}}, solves the equations of \cite{ko1997} to provide the estimated $\Te$ at the end of the collisional region, for a given ion charge-state ratio value. \\

\begin{figure}[t!]
    \centering
    \hspace{-1cm}
    \includegraphics[width = 9cm]{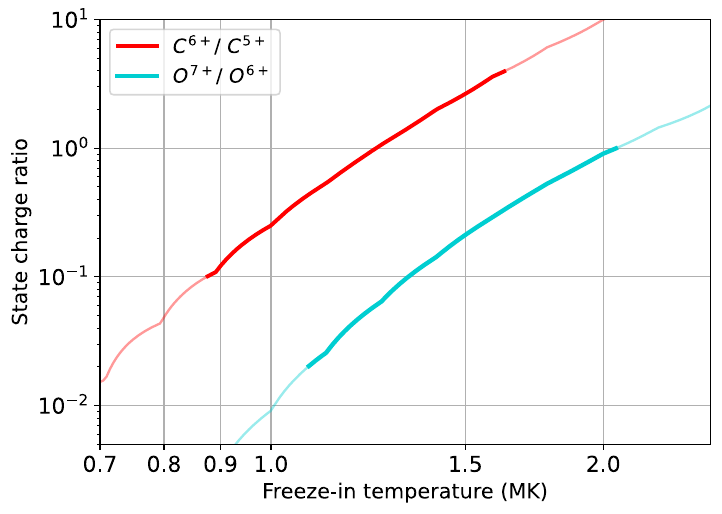}
    \caption{Relationship between the charge-state ratio and the estimated "freeze-in" temperature based on the equations of \cite{ko1997}. 
    The highlighted part of each curve represents the range of charge-state ratio values found in the SolO observations (HIS instrument from 01/17/2022 - 27/04/2023).}
\label{fig_convert_state_charge_Te}
\end{figure}

The carbon and oxygen ions, measured routinely in the solar wind, are typically used to infer $\Te$ in the corona, because their charge state ratios are sensitive to electron temperatures that are encountered in this medium, i.e. from $\sim$~0.5 MK to $\sim$~5 MK for carbon, and $\sim$~1 MK to $\sim$~10 MK for oxygen \citep{Cranmer1999spectroscopic}.
Figure~\ref{fig_convert_state_charge_Te} shows the relationship between the charge-state ratio $\ratioC$ and $\ratioO$ as a function of the freeze-in temperature.

\begin{figure*}[t!]
    %\hspace{-0.8cm}
    \includegraphics[width = 18cm]{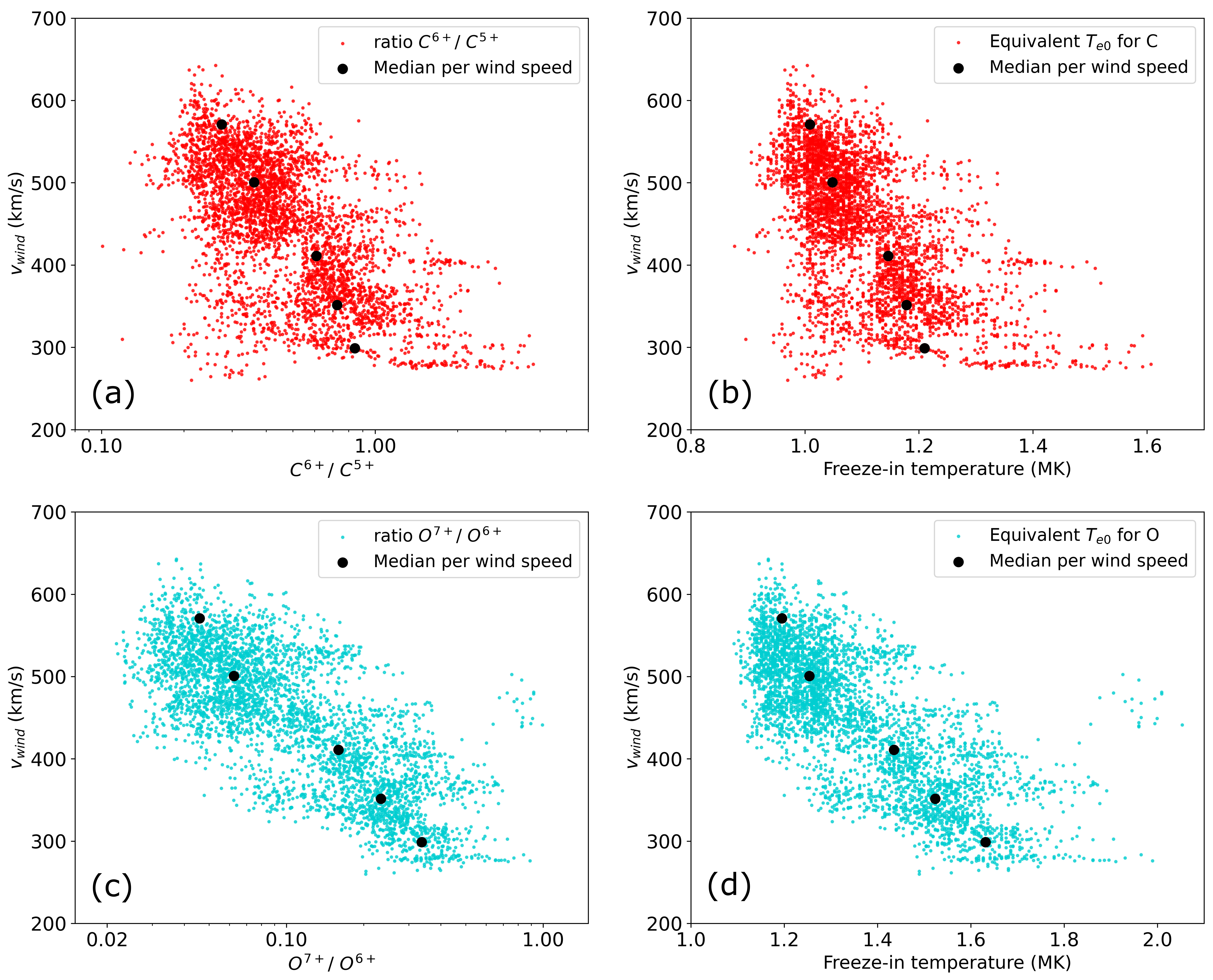}
    \caption{Charge-state ratios and equivalent electron temperature from SolO measurements (from PAS and HIS instruments) between 17/01/2022 and 27/04/2023. Panels (a) and (c) : Observed wind speed as a function of charge-state ratio $\ratioC$ and $\ratioO$, respectively. Panels (b) and (d): Observed wind speed as a function of the equivalent "freeze-in" temperature from panels (a) and (c).}
\label{fig_v_wind_state_charge_equiv_Te}
\end{figure*}

% Using HIS to set state charge limit values
Measurements taken by Solar Orbiter HIS between 17/01/2022 and 27/04/2023 were used to constrain the range of charge-state ratios found in the solar wind. HIS was operating with difficulties before 2022 when the Sun was mostly quiet Consequently, to improve our statistics the data used in the present study were not restricted to the solar minimal activity period considered in \cite{dakeyo2022} and \cite{dakeyo2024b}. 
Consequently, the time period used here is different to that exploited for the original fit of the isopoly populations. Nevertheless, \citet{Wang2016} shows that the ratio $\ratioO$ between minimum and maximum activity levels varies of the order of $\sim$ 35\%, which does not represent a significant change when converted to equivalent coronal  temperatures ($\leq 0.1$ MK) using the scaling shown in Figure~\ref{fig_convert_state_charge_Te}. 

The interval of values sampled by Solar Orbiter is highlighted on each curve of Figure~\ref{fig_convert_state_charge_Te}. The corresponding electron freeze-in temperatures vary between $\sim$ 0.9 MK and 1.6 MK for $\ratioC$, and between $\sim$ 1.1 MK and 2 MK for $\ratioO$. These species do not provide the same freeze-in temperature since they are associated with different $\rf$. Beyond the roughly hydrostatic low corona, the freeze-in height can only be obtained accurately by solving Equation~(2) from \cite{ko1997} for a given radial profile of solar wind speed $u$, density $n$, temperature $\Te$, ionization and recombination rates.

The relevant $\rf$ values are typically below $\sim$ 1.3 $\rs$, so in our wind modeling study, we set the low coronal temperatures to be the same as the estimated freeze-in temperatures (set in the model to $r \approx \rs$).  
% Plot relation Teo with bulk speed from SolO
Moreover, the electron temperature provided by $\ratioO$ and $\ratioC$ can be used directly as the coronal temperature value for isopoly modeling since $\Te$ is assumed constant in isopoly model way above $\rf$ ($\geq$ 2 $\rs$).

In order to model the different wind populations from slow to fast, we need to first determine their corresponding isopoly $\Teo$. To do this, we interpolate the ion charge-state ratios of HIS on the time grid of the proton bulk speed measurements of PAS, and plot $\ratioC$ and $\ratioO$ as a function of wind speed in panels (a) and (c) of the Figure~\ref{fig_v_wind_state_charge_equiv_Te}, respectively. The HIS measurements are then classified in the five wind speed populations of \cite{dakeyo2022, dakeyo2024b}. The associated median values are displayed with black dots. We notice that the carbon observations present a small fraction of data with slow wind observations with low $\ratioC$ and so low coronal $\Te$, that are more representative of fast wind typical value. This could be representative of the Aflvénic slow solar wind \cite{damicis2015}, or it could be due to an instrumental bias since such data points are not as present for $\ratioO$.  
This issue is beyond the scope of the present study. More generally, the results of Figure~\ref{fig_v_wind_state_charge_equiv_Te} agree with previous studies~: the charge state ratio is anti-correlated with the wind speed \citep{Xu2015, Wang2016}.

\begin{table}[t]
\scalebox{1.2}{
\hspace{-1.7cm}
\begin{tabular}{cccccc}
    %%%%%%%%%%%%%%%%%%%%%%%%%%%%%%%%%%%%%%%%%%%%%%%%%%%
    \hline %\hline 
    Wind type & A & B & C & D & E    
     \tabularnewline
    \hline 
    %%%%%%%%%%%%%%%%%%%%%%%%%%%%%%%%%%%%%%%%%%%%%%%%%%%
    Speed (km/s) & 330 & 370 & 410 & 510 & 640
     \tabularnewline
    %\hline 
    %%%%%%%%%%%%%%%%%%%%%%%%%%%%%%%%%%%%%%%%%%%%%%%%%%%
    $\ratioC$ & 0.84 & 0.73 & 0.61 & 0.36 & 0.27
     \tabularnewline
    %\hline 
    %%%%%%%%%%%%%%%%%%%%%%%%%%%%%%%%%%%%%%%%%%%%%%%%%%%
    $\ratioO$ & 0.34 & 0.23 & 0.16 & 0.06 & 0.05
     \tabularnewline
    \hline 
    %%%%%%%%%%%%%%%%%%%%%%%%%%%%%%%%%%%%%%%%%%%%%%%%%%%
    $\TeC$ (MK) & 1.21 & 1.18 & 1.15 & 1.05 & 1.01
     \tabularnewline
    %%%%%%%%%%%%%%%%%%%%%%%%%%%%%%%%%%%%%%%%%%%%%%%%%%%
    $\TeO$ (MK) & 1.63 & 1.52 & 1.44 & 1.25 & 1.20 
     \tabularnewline
    %%%%%%%%%%%%%%%%%%%%%%%%%%%%%%%%%%%%%%%%%%%%%%%%%%%
    \hline
\end{tabular}  
}
\caption{Median charge-state ratio (second and third lines) with their associated equivalent electron temperature in the corona (two bottom lines), computed from HIS and PAS data onboard Solar Orbiter between 17/03/2022 to 17/03/2023. The approximate bulk speed at 1 au, based on the medians profiles in \cite{dakeyo2022} for the populations \textbf{A} to \textbf{E} are displayed on the first line.}
\label{tab_state_charge_Te_med}
\end{table}

The panels (b) and (d) of Figure~\ref{fig_v_wind_state_charge_equiv_Te} show the relation between the coronal $\Te$, derived from Equation~\eqref{eq:condition_ionic_fraction_equilibrium}, with the measured wind speed. The anti-correlations from the two left panels are maintained in the right panels. 
This anti-correlation provides a relation $v$ - $\Teo$ for our isopoly models, which can be used to set the coronal temperatures for each categories wind speed population.
We notice that the temperature $\TeC$ is lower than $\TeO$. Indeed, since the carbon has a lower First Ionization Potential (FIP) than the oxygen, we expect the freeze-in $\TeC$ to be reached lower down in the solar atmosphere than $\TeO$. 
Knowing that collisions are not accounted for in the isopoly model, we consider the highest altitudes as the inner boundaries, to avoid detailed considerations of collisional processes closer to the Sun.

The ionization potential related to the transition from $\text{O}^{6+}$ to $\text{O}^{7+}$ is close to the hydrogen ionization potential, meaning that the freezing ratio $\ratioO$ may represent well an equilibrium of temperature for which $\Te \approx \Tp \approx T_{\text{O}^{7+}}$ in the corona. Thus, we use $\ratioO$ to constrain both protons and electrons with a single freeze-in height \citep{Reames2018}. 
In summary, in the next sections, we set both $\Te$ and $\Tp$ at the coronal base in our solar wind models with the median values from oxygen temperature given in Table \ref{tab_state_charge_Te_med}.

%%%%%%%%%%%%%%%%%%%%%%%%%%%%%%%%%%
%%%%%%%%%%%%%%%%%%%%%%%%%%%%%%%%%%
\subsection{Expansion Factor as Coronal Constraints}
\label{subsec:obs_expansion_factor}
%%%%%%%%%%%%%%%%%%%%%%%%%%%%%%%%%%
%%%%%%%%%%%%%%%%%%%%%%%%%%%%%%%%%%

The magnetic coronal properties in the form of a expansion factor profiles are computed from the same Potential Field Source Surface (PFSS) extrapolations of ADAPT magnetograms used in the study of \cite{dakeyo2024b}. The PFSS reconstruction provides the local 3D magnetic field components from $\rs$  until the source surface located at $2.5 \: \rs$. Based on the conservation of magnetic flux, the expansion factor radial evolution is expressed as~:  
  \begin{align}
    f(r) = \frac{ \Br(\rs)}{\Br(r)} \frac{\rs^2}{r^2}
  \end{align}
where $\Br(r)$ is the magnetic field radial component at the radial position $r$ and $\rs$ is the Sun radius. 

The expansion factor profiles $f(r)$ were computed to perform a magnetic connectivity study in\cite{dakeyo2024b}, by applying a sunward backmapping from Solar Orbiter to the solar photosphere \citep{Rouillard2020b}. The period covered by the $f(r)$ profiles goes from 01/08/2020 to 17/03/2022 during the solar minima period. As in \cite{dakeyo2024b}, we classify our model solutions according to the maximum (final) expansion factor values $f(\rss) = \fss$, into six $\fss$ bins: $<$ 7, [7, 20], [20, 50], [50, 100], [100, 250], and $>$ 250. However, in the present study we only consider the smallest bin for $\fss<7$ and the previous last bin $100 < \fss< 250$ to represent the extrema flux tube characteristics. 
See \cite{dakeyo2024b} for more details.

%%%%%%%%%%%%%%%%%%%%%%%%%%%%%%%%%%
%%%%%%%%%%%%%%%%%%%%%%%%%%%%%%%%%%
\section{Isopoly Modeling Limitations: Charge-State Ratio}
\label{sec:isopoly_state_charge}
%%%%%%%%%%%%%%%%%%%%%%%%%%%%%%%%%%
%%%%%%%%%%%%%%%%%%%%%%%%%%%%%%%%%%

\begin{figure*}[t!]
    \includegraphics[width = 18.cm]{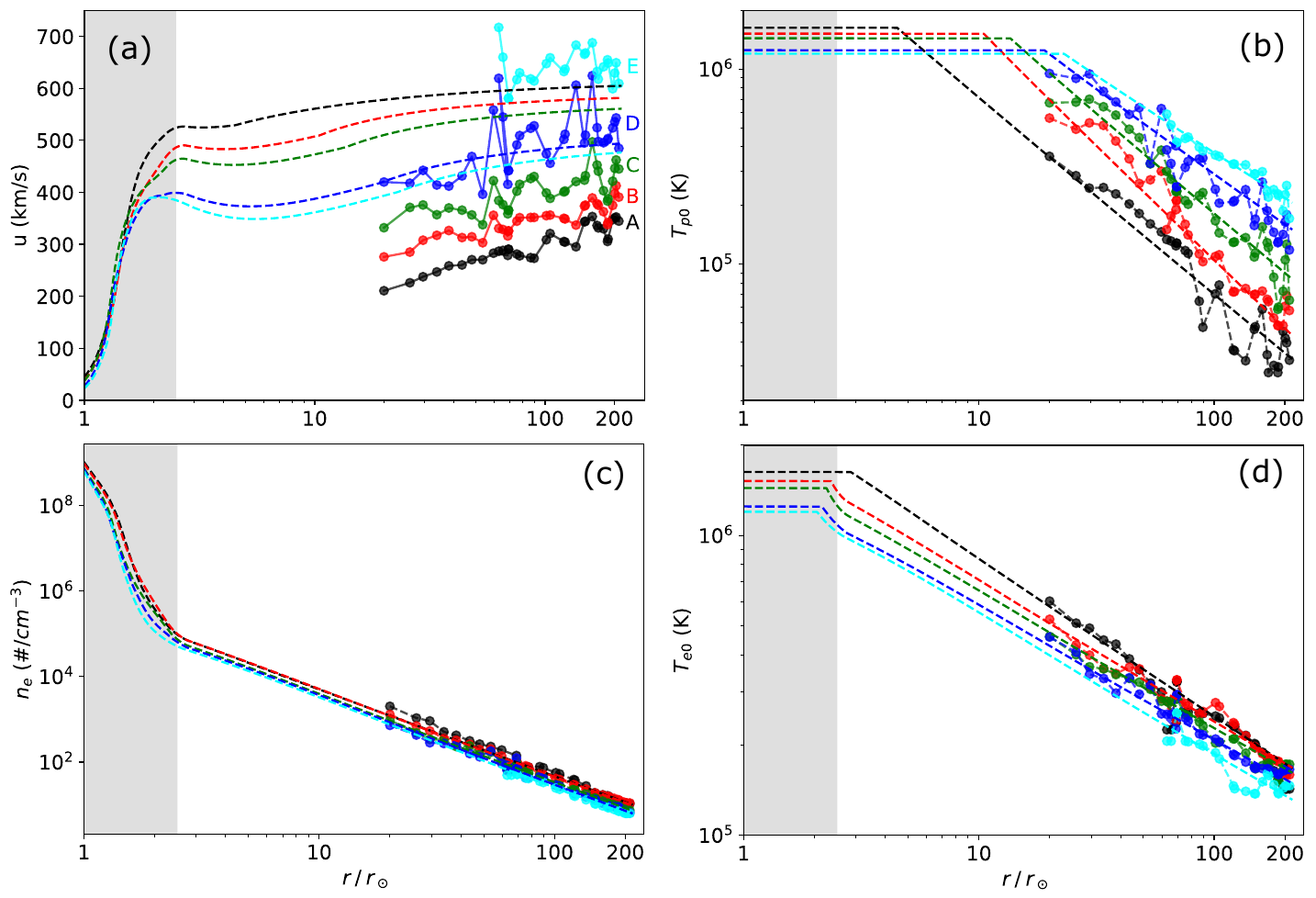}
    \caption{Isopoly models computed from Equation~\eqref{eq_momentum_all_term_detailled_fexp} associated with expansion factor profiles computed from PFSS, fitted to the data set used by \cite{dakeyo2022} and the $\Teo$ derived from the Solar Orbiter oxygen charge-state ratios. The f-supersonic solutions are shown in dashed lines. No f-subsonic solutions exist for the chosen set of input parameters. 
    Five isopoly models are computed for the expansion factor bin $ 100 < \fss <250$ \citep[as defined in][]{dakeyo2024b}, no wind solutions exist for the bin $\fss <7$ with their given input parameters. Panel (a)~: Velocity profiles; Panel (b)~: Proton temperature profiles; Panel (c)~: Density profiles; Panel (d): Electron temperature profiles. 
    The data used for the fit are added in panels (a), (b), (c) and (d) as dots connected by straight segments of the same color of the corresponding wind population. 
    The region of super radial expansion (up to $\rss=2.5~\rs$) is delimited by the gray shaded area.}
\label{fig_u(r)_Tp(r)_ne(r)_Te(r)_mapped_data_f-subsup_Tp_Te_modif}
\end{figure*}

% Fit Teo anti-correlated for isopoly
In order to incorporate the source temperature constraints given by the charge-state ratio in the isopoly models, we explore the ability of the isopoly model to reproduce at the same time the anti-correlation of the bulk velocity with the coronal temperature $\Tpo$ and $\Teo$ (deduced in Section \ref{sec:isopoly_state_charge}), and the interplanetary observations while accounting for the typical coronal expansion factor profiles. We use the isopoly models presented in \cite{dakeyo2024b}, setting $\Tpo = \Teo = \TeO$ (deduced from $\ratioO$) and fitting the interplanetary medium observations of Parker Solar Probe and Helios.

\noindent
% Description of the main equations
The considered isopoly equations from \cite{dakeyo2024b} with super expansion, are set with the conservation of momentum~: 

\begin{equation}
   n\, m_p\, u \frac{d u}{d r} = - \sums \frac{d \Ps}{d r}  - n\, m_p \frac{G \, M}{r^2},
  \label{eq_momentum_sans_hypothese}
\end{equation}

with $n$ the density, $\Ps$ the plasma pressure, $m_p$ the proton mass, $M$ the Sun mass, and where the sum over the species $s$ is taken over protons ($\mathrm p$)  and electrons ($\mathrm e$). The temperature is
\begin{align}
      \Ts(r) = \Tso \: \bigg( \frac{n(r)}{n(\risos)} \bigg)^{\gams -1} 
      \label{eq_Ts_isopoly}
      \mbox{with~} %\rightarrow
    \left\{
        \begin{array}{ll}
             r\: \leq \: \risos: \quad \gamma_s = 1     \\
            r\: > \: \risos: \quad \gamma_s > 1 
        \end{array}
    \right.
\end{align}
where $\risos$ is the distance below which the expansion is isothermal, and $\gams$ is the polytropic index. The density can be expressed using mass flux conservation, where $n \: u \: f \: r^2 = C$, and $C$ is a constant determined from observations. The resolution of Equation \eqref{eq_momentum_sans_hypothese} including $f(r)$ follows the same development as in \cite{dakeyo2022}, with an additional term in the derivative of the density: 
\begin{align}
   \frac{d \ntils}{d r} = - \frac{1}{n(\risos)} \frac{C}{f r^2} \bigg[ \frac{2}{ur} + \frac{1}{u^2} \frac{du}{dr} + \frac{1}{uf} \frac{df}{dr} \bigg],
\end{align}

\noindent
where $\ntil = n(r)/n(\risos)$. Its inclusion in the momentum Equation \eqref{eq_momentum_sans_hypothese} leads to:
\begin{align}
    \frac{du}{dr} 
     \bigg[ 1 -  \frac{c^2}{u^2} \bigg]  
     = \frac{1}{u r} \bigg[  c^2 \:  
        %\bigg(2 - f r \: \frac{d (1/f)}{dr} \bigg) 
        \bigg(1 + \frac{r}{2} \: \frac{d \log f}{dr} \bigg)
        - \frac{G \, M}{r} \bigg],
\label{eq_momentum_all_term_detailled_fexp}
\end{align}
where $c^2=\sums \cs^2 \xs$, $\: \xs= \ntil^{\: \gams -1} $. 
Equation~\eqref{eq_momentum_all_term_detailled_fexp} is similar to the isopoly Equation (B8) presented in \cite{dakeyo2022}, with an extra term related to $f(r)$ expressing the effect of the super radial expansion. 

The existence of transonic wind solution for f-subsonic and f-supersonic solution type (respectively subsonic and partly supersonic in the super expansion region), is fully detailed in \cite{dakeyo2024b}. The transonic criterion requires the validity of  Equation~\eqref{eq_momentum_all_term_detailled_fexp} for $du/dr \neq 0$ at each radial distance. The critical radius $\rc$ is then derived by solving  Equation~\eqref{eq_momentum_all_term_detailled_fexp} for $u = c$. The main difference between the f-subsonic and f-supersonic solution is a change in the location of the critical radius $\rc$, which is either $\geq 3 \ \rs$ or $\leq 2 \ \rs$ respectively, regardless of the final wind speed. The the sonic point in the f-supersonic solutions occurs in the region of super-radial expansion, increasing the efficiency of the work of pressure force in the near Sun region.
A Python code solving the above isopoly solutions is available at \hyperref[https://github.com/jbdakeyo/IsopolySolarWind]{\url{https://github.com/jbdakeyo/IsopolySolarWind}}.
For further information on isopoly equations, we refer to the above link and to \cite{dakeyo2022} and \citep{dakeyo2024b}.

Using coronal temperature from Table~\ref{tab_state_charge_Te_med}, the fitting of the isopoly models to their respective wind populations is done manually, modifying $\risop$ and $\risoe$ to visually match the observed $\Tp$ and $\Te$ profiles. 
Since $\Tpo$ and $\Teo$ are fixed, the resulting isopoly speed profiles are directly derived fitting proton and electron temperature profiles.\\

A summary of the solar wind speed parameters is shown in Appendix \ref{sec:appendix_isopoly_param_table} in Table \ref{tab_updated_isopoly_param_modif_Te0}. The corresponding curves are shown in Figure~\ref{fig_u(r)_Tp(r)_ne(r)_Te(r)_mapped_data_f-subsup_Tp_Te_modif}.
% Wind profiles description
Panel (a) shows that none of the wind populations, except the \textbf{D} wind, matches its corresponding wind speed for the imposed coronal and interplanetary conditions. Moreover, the wind speed populations show a reverse order compared to their expected 1 au speed, implying that the observed ordering of coronal temperature for isopoly modeling globally sets the ordering of wind speed at 1 au. This causes the supposed slow and intermediate wind populations \textbf{A}, \textbf{B} and \textbf{C} to carry too much heating with a large increase in bulk velocity, while the supposed fast wind population \textbf{E} is not heated enough. All of this results in a 1 au bulk velocity inversion compared to expectations. 
Furthermore, we see that for $ 100 < \fss <250$ only f-supersonic solutions exist with the isopoly modeling. 
Indeed, the numerical resolution of the Equation~\eqref{eq_momentum_all_term_detailled_fexp} locates almost for all models the $\rc$ values in the super expansion region with $\rc < 2.5 \rs$ (as shown in Table \ref{tab_updated_isopoly_param_modif_Te0}). In contrast, in the case of small expansion factor values ($\fss <$7), when computing u(r) anti-sunward from $\rc$, the input parameters leads the modeled wind speed to go down the sound speed (breeze solution). Therefore the wind solution can be neither f-subsonic nor f-supersonic, so the input conditions cannot be solved accordingly to a transonic solution.

All these results indicate that the isopoly approach cannot incorporate the coronal information carried by the charge-state ratio observations. This suggests that the isothermal assumption in the corona is too restrictive to model the coronal and interplanetary observations with such setting. This calls for further improvements of the model.

%%%%%%%%%%%%%%%%%%%%%%%%%%%%%%%%%%
%%%%%%%%%%%%%%%%%%%%%%%%%%%%%%%%%%
\section{Generalized Two Thermal Regime~: Bipoly Modeling}
\label{sec:two_thermal_regime_model}
%%%%%%%%%%%%%%%%%%%%%%%%%%%%%%%%%%
%%%%%%%%%%%%%%%%%%%%%%%%%%%%%%%%%%

%%%%%%%%%%%%%%%%%%%%%%%%%%%%%%%%%%
\subsection{Bipoly Equation Formalism and Solving Method}
\label{subsec:bipoly_equation_formalism}
%%%%%%%%%%%%%%%%%%%%%%%%%%%%%%%%%%

% How to overcome isopoly limitations
As presented in the previous section, the inclusion of the charge-state ratio in the isopoly approach from \cite{dakeyo2024b} implies that the isopoly model struggles to reproduce both coronal and interplanetary properties. In the present section we investigate an improvement of the isopoly approach, that we called the "bipoly" model. The fundamental assumption of splitting the wind evolution into two distinct thermal regimes is kept. However, the isothermal condition $\gamma=1$ near the Sun being too restrictive, we relaxed this condition by considering the first thermal regime as another polytropic regime with $\gamma \neq 1$. 
The polytropic closure between temperature $\Ts$ and density $n$ is defined as~:
\begin{align}
     \Ts(r) = \Tsc \: \bigg( \frac{n(r)}{n(\rc)} \bigg)^{\gams -1} 
      \rightarrow
    \left\{
        \begin{array}{ll}
            r\: \leq \: \rpolys: \quad \gams = \gamso    \\
            r\: > \: \rpolys: \quad \gams = \gamsl 
        \end{array}
    \right.,
\end{align}
where $\gamso >0$ is expected to be close to 1, while $\gamsl > 1$.
% Meaning of gamma < 1 in bipoly modeling 
The values of $\gams$ in the first thermal regime can be set either $>$1 or $<$1. 
A value of $\gamma <1$ (sub-isothermal) may be surprising since already $\gamma =1$ applied to all distances implies an infinite input of thermal energy into the system from an heating source. %Then, $\gamma <1$ may seem difficult to conceive physically. 
However, here $\gamma <1$ is only permitted in a finite volume, so the energy input stays finite. The possibility of $\gamma <1$ comes from the fact that we want to model an increase of the plasma temperature. Considering $T \propto r^{-\alpha}$ and $n \propto r^{-\beta}$, identifying the exponent in the polytropic relation leads to~:
\begin{align}
    T \propto n^{\gamma-1}
    \quad \Rightarrow \quad
    r^{-\alpha} \propto r^{-\beta (\gamma-1)}
    \quad \Rightarrow \quad
    \gamma = \frac{\alpha + \beta}{\beta}.
\end{align}
An increase of temperature is set by the exponent $\alpha <0$, so $\gamma<1$.
In terms of plasma thermodynamics, during its evolution the system receives a larger amount of thermal energy than it losses with radial evolution, causing the system temperature to rise.

% Introduction of bipoly specific parameters
The bipoly equations can be set up similarly to the set of Equations~\eqref{eq_Ts_isopoly}-\eqref{eq_momentum_all_term_detailled_fexp},  but with a different way to solve them and a relaxation of the $\gamma=1$ condition.
We define each thermal regime separately, and impose the continuity of the plasma parameters at the boundaries ($r=\rpolys$). This requires the definition of each region, as well as a critical radius $\rc$ that characterizes the evolution of a given region.

Solving Parker's like equations with an explicit scheme requires starting from the critical radius $\rc$, and then computing the momentum equations sunward and outward to derive the full wind transonic wind solution. The definition of the sonic point is critical since it defines the reference quantities $\cs$ and $\rc$ that are used in Equation~\eqref{eq_momentum_all_term_detailled_fexp} to obtain the speed at all distances. However, in the case of the casting of several thermal regimes, the information of the reference point must be used accordingly to the critical radius and sound speed that guarantee a transonic evolution for the local polytropic law, i.e. accordingly to the local values of $\gamp$ and $\game$. 
Therefore, to guarantee transonic solution at all distances, we must first find the reference critical radius $\rc$ and sound speed $\cs$ in each region. The different regions are set by the two values of $\gams$ taken for both protons and electrons (s=\{p,e\}), $\gamso$ below $\rpolys$ (the distance that set the end of the first thermal regime), and $\gamsl$ above $\rpolys$. Considering a two-fluid modeling, as in the previous isopoly modeling, the distance $\rpolyp$ and $\rpolye$ are allowed to be different, consequently there also exists a region between $\rpolyp$ and $\rpolye$ with different reference critical radius and sound speed. 
To summarize, the bipoly solution considers three thermal regions with for each a different set of $\{\rc,\cs\}$ reference values, however it results in only a single sound speed crossing and a unique pair of $\{\rc,\cs\}$ values representing this actual sound speed crossing. \\

\begin{figure*}[t]
    \includegraphics[width = 18.cm]{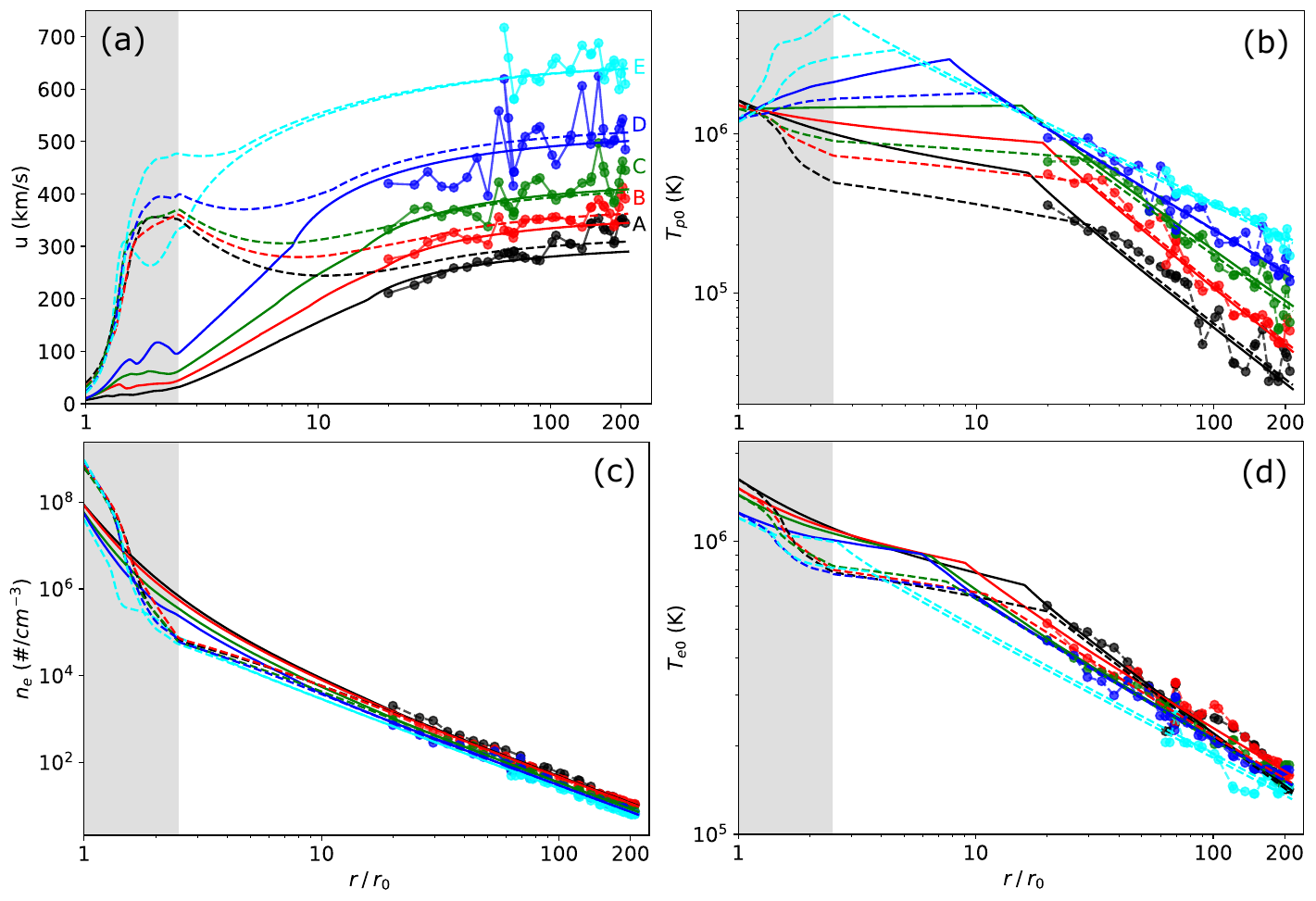}
    \caption{Same as Figure~\ref{fig_u(r)_Tp(r)_ne(r)_Te(r)_mapped_data_f-subsup_Tp_Te_modif} but with bipoly solutions computed from Equation~\eqref{eq_momentum_all_term_detailled_fexp} fitted to the in-situ data set used by \cite{dakeyo2022}. Models include the flux tube expansion factor profiles computed from PFSS, and $\Teo$ deduced from SolO charge-state ratio. The f-subsonic and the f-supersonic solutions are plotted in solid and dashed lines respectively. Five isopoly models are computed for the two bins $\fss < 7$ and $ 100 < \fss <250$. Panel (a)~: Velocity profiles; Panel (b)~: Proton temperature profiles; Panel (c)~: Density profiles; Panel (d): Electron temperature profiles. The data used for fitting are added in all panels as dots linked with straight segments of same color of its corresponding wind population. The region of super radial expansion (up to $\rss=2.5~\rs$) is delineated by the gray shaded area.
    }
\label{fig_bipoly_u(r)_Tp(r)_ne(r)_Te(r)_mapped_data_f-subsup_Tp_Te_modif}
\end{figure*}

\begin{table*}[t]
\hspace{1.5cm}
\scalebox{1}{
%\hspace{-1.7cm}
\begin{tabular}{cccccc}
    \multicolumn{6}{c}{Bipoly typical parameters for $\fss <$  7}
     \tabularnewline
    %%%%%%%%%%%%%%%%%%%%%%%%%%%%%%%%%%%%%%%%%%%%%%%%%%%
    \hline \hline
    Typical speed & Slow & Medium - Slow & Medium & Medium - Fast & Fast    
     \tabularnewline
    \hline \hline
    %%%%%%%%%%%%%%%%%%%%%%%%%%%%%%%%%%%%%%%%%%%%%%%%%%%
    %\hline %\hline 
    Wind type & A & B & C & D & E     
     \tabularnewline
    \hline 
    %%%%%%%%%%%%%%%%%%%%%%%%%%%%%%%%%%%%%%%%%%%%%%%%%%%
    $\Tpc$ (MK) & 0.720 & 1.035  & 1.490 & 2.410 & 2.305 
     \tabularnewline
    %\hline 
    %%%%%%%%%%%%%%%%%%%%%%%%%%%%%%%%%%%%%%%%%%%%%%%%%%%
    $\Tec$ (MK) & 0.850  & 0.920 & 0.940 & 0.960 & 1.110
     \tabularnewline
    %\hline 
    %%%%%%%%%%%%%%%%%%%%%%%%%%%%%%%%%%%%%%%%%%%%%%%%%%%
    $\rpolyp$ ($\rs$) & 16.5 & 19.0 & 15.5 & 7.7 & 2.7 
     \tabularnewline
    %\hline 
    %%%%%%%%%%%%%%%%%%%%%%%%%%%%%%%%%%%%%%%%%%%%%%%%%%%
    $\rpolye$ ($\rs$) & 16.0 & 9.0 & 6.5 & 5.9 & 2.6
     \tabularnewline
    %\hline 
    %%%%%%%%%%%%%%%%%%%%%%%%%%%%%%%%%%%%%%%%%%%%%%%%%%%
    $\gampo$ & \textbf{1.100} & \textbf{1.050} & \textbf{0.995} & \textbf{0.902} & \textbf{0.757}
     \tabularnewline
    %\hline 
    %%%%%%%%%%%%%%%%%%%%%%%%%%%%%%%%%%%%%%%%%%%%%%%%%%%
    $\gameo$ & \textbf{1.080} & \textbf{1.065} & \textbf{1.060} & \textbf{1.040} & \textbf{1.030}    
    \tabularnewline
    \hline    
    %%%%%%%%%%%%%%%%%%%%%%%%%%%%%%%%%%%%%%%%%%%%%%%%%%%
    $\Tpo$ (MK) & 1.63 & 1.52 & 1.44 & 1.25 & 1.20
     \tabularnewline
    %\hline 
    %%%%%%%%%%%%%%%%%%%%%%%%%%%%%%%%%%%%%%%%%%%%%%%%%%%
    $\Teo$ (MK) & 1.63 & 1.52 & 1.44 & 1.25 & 1.20
    \tabularnewline
    %\hline
    %%%%%%%%%%%%%%%%%%%%%%%%%%%%%%%%%%%%%%%%%%%%%%%%%%%
    $\Tpmax$ (MK) & \textbf{1.63} & \textbf{1.52} & \textbf{1.51} & \textbf{2.97} & \textbf{5.74}
     \tabularnewline
    %\hline 
    %%%%%%%%%%%%%%%%%%%%%%%%%%%%%%%%%%%%%%%%%%%%%%%%%%%
    $u_{0}$ (km/s) & 7 & 11 & 10 & 10 & 25 
     \tabularnewline
    %\hline  
    %%%%%%%%%%%%%%%%%%%%%%%%%%%%%%%%%%%%%%%%%%%%%%%%%%%
    $u_{1 au}$ (km/s) & \textbf{290} & \textbf{345} & \textbf{409} & \textbf{501} & \textbf{639}  
     \tabularnewline
    %%%%%%%%%%%%%%%%%%%%%%%%%%%%%%%%%%%%%%%%%%%%%%%%%%%
    $n_{0}$ ($10^7 \# / cm^{-3}$) & 9 & 8 & 6 & 5 & 4
     \tabularnewline
    %\hline
    %%%%%%%%%%%%%%%%%%%%%%%%%%%%%%%%%%%%%%%%%%%%%%%%%%%
    $r_c$ ($\rs$) & 6.75 & 5.58  & 4.65 & 3.64 & 1.30
     \tabularnewline 
    \hline 
    \tabularnewline

    \multicolumn{6}{c}{Bipoly typical parameters for $100 < \fss < 250$  }
     \tabularnewline
    %%%%%%%%%%%%%%%%%%%%%%%%%%%%%%%%%%%%%%%%%%%%%%%%%%%
    \hline 
    Wind type & A & B & C & D & E     
    \tabularnewline
    \hline 
    %%%%%%%%%%%%%%%%%%%%%%%%%%%%%%%%%%%%%%%%%%%%%%%%%%%
    $\Tpc$ (MK) & 1.240 & 1.250 & 1.280 & 1.360 & 1.575
     \tabularnewline
    %\hline 
    %%%%%%%%%%%%%%%%%%%%%%%%%%%%%%%%%%%%%%%%%%%%%%%%%%%
    $\Tec$ (MK) & 1.380 & 1.280 & 1.250 & 1.090 & 1.070
     \tabularnewline
    %\hline 
    %%%%%%%%%%%%%%%%%%%%%%%%%%%%%%%%%%%%%%%%%%%%%%%%%%%
    $\rpolyp$ ($\rs$) & 30.0 & 30.0 & 27.0 & 11.5 & 4.5
     \tabularnewline
    %\hline 
    %%%%%%%%%%%%%%%%%%%%%%%%%%%%%%%%%%%%%%%%%%%%%%%%%%%
    $\rpolye$ ($\rs$) & 20.0 & 11.0 & 7.5 & 8.8 & 3.5
     \tabularnewline
    %\hline 
    %%%%%%%%%%%%%%%%%%%%%%%%%%%%%%%%%%%%%%%%%%%%%%%%%%%
    $\gampo$ & \textbf{1.130} & \textbf{1.080} & \textbf{1.050} & \textbf{0.970} & \textbf{0.905}
     \tabularnewline
    %\hline 
    %%%%%%%%%%%%%%%%%%%%%%%%%%%%%%%%%%%%%%%%%%%%%%%%%%%
    $\gameo$ & \textbf{1.080} & \textbf{1.070} & \textbf{1.060} & \textbf{1.050} & \textbf{1.040}
     \tabularnewline
    \hline
    %%%%%%%%%%%%%%%%%%%%%%%%%%%%%%%%%%%%%%%%%%%%%%%%%%%
    $\Tpo$ (MK) & 1.63 & 1.52 & 1.44 & 1.25 & 1.20
     \tabularnewline
    %\hline 
    %%%%%%%%%%%%%%%%%%%%%%%%%%%%%%%%%%%%%%%%%%%%%%%%%%%
    $\Teo$ (MK) & 1.63 & 1.52 & 1.44 & 1.25 & 1.20
    \tabularnewline
    %\hline
    %%%%%%%%%%%%%%%%%%%%%%%%%%%%%%%%%%%%%%%%%%%%%%%%%%%
    $\Tpmax$ (MK) & \textbf{1.63} & \textbf{1.52} & \textbf{1.44} & \textbf{1.83} & \textbf{3.39}
     \tabularnewline
    %\hline 
    %%%%%%%%%%%%%%%%%%%%%%%%%%%%%%%%%%%%%%%%%%%%%%%%%%%
    $u_{0}$ (km/s) & 40  & 34 & 32 & 23 & 21
     \tabularnewline
    %\hline  
    %%%%%%%%%%%%%%%%%%%%%%%%%%%%%%%%%%%%%%%%%%%%%%%%%%%
    $u_{1 au}$ (km/s) & \textbf{309} & \textbf{362} & \textbf{403} & \textbf{517} & \textbf{639}  
     \tabularnewline
    %%%%%%%%%%%%%%%%%%%%%%%%%%%%%%%%%%%%%%%%%%%%%%%%%%%
    $n_{0}$ ($10^7 \# / cm^{-3}$) & 61 & 70 & 64 & 89 & 95
     \tabularnewline
    %\hline
    %%%%%%%%%%%%%%%%%%%%%%%%%%%%%%%%%%%%%%%%%%%%%%%%%%%
    $r_c$ ($\rs$) & 1.31 & 1.36 & 1.31 & 1.32 & 1.34 
     \tabularnewline
    \hline 
\end{tabular} 
}
\caption{Bipoly input and output parameters associated with the bipoly curves considering the expansion factor modeling, $\Tpo$ and $\Teo$ are based on charge-state ratio measurements of Figure~\ref{fig_v_wind_state_charge_equiv_Te}. The top panel shows parameters associated with low expansion $\fss<7$ and the bottom panel with high expansion $100 < \fss < 250$. The bipoly free input parameters $\Tpc$, $\Tec$, $\rpolyp$, $\rpolye$, $\gampo$, $\gameo$ are on the six top lines. The input parameters have been manually modified for bipoly models to fit coronal constraints provided by $\TeO$, and interplanetary observations of $u$, $\Tp$, $\Te$, $n$. For all models, the interplanetary polytropic indices $\gampl$ and $\gamel$ are observationally constrained, and are the same as in \cite{dakeyo2022} 
$\gampl = (1.57, 1.59, 1.52, 1.44, 1.35)$ and $\gamel = (1.29, 1.24, 1.23, 1.23, 1.21)$.
The bipoly output parameters (coronal temperatures $\Tpo$, $\Teo$, maximum proton temperature $\Tpmax$, coronal and 1~au velocity $u_0$, $u_{1au}$, coronal density $n_0$, and critical (sonic) radius $\rc$) are shown on the 6 bottom lines. 
The features of most interest are shown in bold.
 }
\label{tab_bipoly_param_modif_Te0}
\end{table*}

The equation of dynamic is similar as Equation~\eqref{eq_momentum_all_term_detailled_fexp}, but the constant used to set the polytropic relation between $\Ps$ and $n$ is no longer taken at $\risos$, but at $\rc$, the sonic point~:
\begin{align}
    \Ps = \Ps(\rc) \: \bigg( \frac{n}{n(\rc)} \bigg)^{\gams} 
    = \Ps(\rc) \: \ntil^{\gams},
    \label{eq_Ps_bipoly}
\end{align}
where $\rc$ is set to a different value in each of the three thermal evolution regions dependent of polytropic indexes combination. 
We must note that Equation \eqref{eq_Ps_bipoly} is only valid for a single value of gamma, and is used in the momentum equation \eqref{eq_momentum_sans_hypothese} as the local polytropic relation to follow. Indeed, the constants $\Ps(\rc)$ and $n(\rc)$ are taken locally to represent the local critical radius i.e. the local thermal regime, without which $\Ps$ would be discontinuous. Consequently, this expression cannot be used to cast multiple thermal regimes. The actual calculation of $\Ps$ through all the different thermal regimes is taken from the perfect gas equations~:
\begin{align}
    \Ps = n k_B\Ts
\end{align}
where $k_B$ the Boltzmann constant. Since $\Ts$ and $n$ are continuous by construction, Ps is then also continuous.

% Bipoly modeling fitting observations and Teo from state charge
To illustrate the benefit of relaxing the isopoly isothermal condition into a polytropic one with 
the bipoly model, we recalculate the solutions shown in Figure~\ref{fig_u(r)_Tp(r)_ne(r)_Te(r)_mapped_data_f-subsup_Tp_Te_modif} with the bipoly modeling, which use the coronal temperature $\Tpo$ and $\Teo$ deduced from the charge-state ratio observations. We must note that the $\rf$ values have not been determined here. This would require a complete radial resolution of the ionization equations, but since $\rf$ values are estimated to be below $1.3 \: \rs$ \citep{ko1997}, and since we only need an estimate of $\Tpo$ and $\Teo$, for simplicity we use $\TeO$ as coronal temperature target for bipoly models at $r=\rs$. 
The main features and free input parameters of the models for small and large expansion factor bins defined in Section~\ref{subsec:obs_expansion_factor}, are summarized in the top and bottom panels of Table~\ref{tab_bipoly_param_modif_Te0}, respectively.

%%%%%%%%%%%%%%%%%%%%%%%%%%%%%%%%%%
\subsection{Bipoly Solar Wind Modeling}
\label{subsec:bipoly_solar_wind_modeling}
%%%%%%%%%%%%%%%%%%%%%%%%%%%%%%%%%%

% Speed profiles description
The speed profiles on panel (a) of Figure~\ref{fig_bipoly_u(r)_Tp(r)_ne(r)_Te(r)_mapped_data_f-subsup_Tp_Te_modif} show that all wind populations can be better modeled with bipoly models than with isopoly ones (Figure~\ref{fig_u(r)_Tp(r)_ne(r)_Te(r)_mapped_data_f-subsup_Tp_Te_modif}), with the inclusion of the anti-correlation asymptotic wind speed - coronal temperature. We compute the two expansion factor cases as described in Section~\ref{subsec:obs_expansion_factor}, with $\fss < 7$ and $ 100 < \fss <250$, for each wind population. The large $\fss$ models present exclusively f-supersonic solutions (dashed lines) and they lead to the faster wind at large solar distance. 
The small $\fss$ models present f-subsonic solutions for \textbf{A}-\textbf{D} populations (solid lines). The solution for population \textbf{E} is similar to the solutions for large $\fss$ models (except in the region $1.5\,\rs < r < 5\,\rs $).

For small $\fss$ values we distinguish the evolution of each wind speed profile for nearly all $r$ values. In contrast, in the super expansion region (grey shaded region), all wind populations for large $\fss$ (dashed lines) show similar velocities and acceleration rate until they reach $\sim$ 300 km/s at $\sim$ 1.5 $\rs$ (in sharp contrast with Figure~\ref{fig_u(r)_Tp(r)_ne(r)_Te(r)_mapped_data_f-subsup_Tp_Te_modif}). At larger distances, they each evolve differently to reach different asymptotic wind speeds. In the super expansion region the work of the pressure force to convert thermal energy into kinetic energy is comparable, regardless of the wind temperature evolution close to the Sun (increasing or decreasing as shown in panels (b,d)). Thus, the super expansion shapes the plasma acceleration more than the local thermodynamic conditions of the plasma ($T$ and $n$). In addition, the wind \textbf{A} solution reaches a maximum wind speed around 2.5 $\rs$ comparable to 1~au values, which is not commonly expected regarding that f-subsonic solutions is more generally considered in the state of the art 
\citep[][and references therein]{dakeyo2024b}.

% Tp profiles description
Regarding $\Tp$ profiles in panel (b) we recall that all the coronal values satisfy the charge-state ratio conditions at $r= \rs$. In the case where the first thermal regime requires $\gampo<1$, the maximum proton temperature is reached at the distance $\rpolyp$ where the thermal regime changes, while in the case of $\gampo\geq1$, it is reached at $\rs$. The maximum values are of the order of $\sim$ 1.5 MK for slow and medium winds (populations \textbf{A} to \textbf{C}), while it goes up to 5.7 MK at 2.7 $\rs$ for the fastest wind (populations \textbf{E}). These temperatures are in agreement with the hydrogen temperature determined by \cite{Cranmer2002coronal} from spectroscopic observations (expected to be similar to the proton temperature), which is estimated to be between 1 MK and 6 MK between 1 $\rs$ and 4 $\rs$.

% Density profiles description
The density profiles in panel (c) show higher coronal values $n_0$, for larger $\fss$ compared to small $\fss$ curves.  For the larger $\fss$, the $n(r)$ profiles have a similar behavior to panel (c) of Figure ~\ref{fig_u(r)_Tp(r)_ne(r)_Te(r)_mapped_data_f-subsup_Tp_Te_modif} with a sharp decrease of density in the corona and high values at the coronal base.
Globally we obtain lower coronal densities than the isopoly solutions shown in \cite{dakeyo2024b}, here of the order of $n_0 = 10^7 - 10^9$ $\#.cm^{-3}$. This makes the bipoly densities more consistent with the typical $\sim 10^8$ $\#.cm^{-3}$ deduced from spectroscopic observations \citep{bemporad2017}. 

% Te profiles description
Panel (d) shows the electron temperature profiles that also satisfy the observed freeze-in temperatures at the coronal base. However, in contrast to $\Tp$ profiles, the $\Te$ profiles are all fitted with $\game > 1$ to match both the coronal and interplanetary conditions. Moreover we find that larger $\fss$ values induce larger variations in the temperature profiles for both $\Tp$ and $\Te$. Indeed, the conservation of the mass flux with the polytropic relation induces a stronger temperature decrease or increase in the region of super-radial expansion. 

%%%%%%%%%%%%%%%%%%%%%%%%%%%%%%%%%%
%%%%%%%%%%%%%%%%%%%%%%%%%%%%%%%%%%
\section{Conclusion}
\label{sec:conclusion}
%%%%%%%%%%%%%%%%%%%%%%%%%%%%%%%%%%
%%%%%%%%%%%%%%%%%%%%%%%%%%%%%%%%%%
% Resume of the study
This paper investigates the implications of adding  observed coronal temperature as a new set of constraints to our isopoly wind model\citep{dakeyo2022, dakeyo2024b}. For that, we use charge-state ratio measurements from Solar Orbiter from 17/01/2022 to 27/04/2023. The charge state of oxygen $\ratioO$ from Solar Orbiter and the assumptions of the "freeze in" temperature in the low corona \citep{ko1997}, provide estimated coronal temperatures as new constraints at the base of the solar wind for both protons and electrons.

We show that the isopoly modeling cannot reproduce simultaneously the interplanetary properties of solar wind populations (from slow to fast wind) and the observed coronal wind properties. Indeed, the isothermal assumptions appears to be too strong to capture coronal properties with more precision than order of magnitude of the coronal plasma parameters. 

% Conclusion bipoly in comparison of isopoly
To alleviate the isothermal assumption, we relax the constraint of an isothermal plasma to a polytropic one for the first thermal regime of the solar wind evolution (close to the Sun), leading to what we called bipoly model. The bipoly model fits successfully the interplanetary observations ($u$, $\Tp$, $\Te$ and $n$) from Helios and Parker Solar Probe, and according to $\Tpo$ and $\Teo$ calculated from the charge-state ratio. This improvement considerably reduces the required coronal temperature values $\Tpo$ to model fast winds compared to \cite{dakeyo2024b}. Moreover, the bipoly coronal velocities $u_0$ and densities $n_0$, agree better with the typical values $\lesssim$ 100 km/s and $\sim \: 10^8 \: \#.cm^{-3}$ respectively, derived from remote-sensing measurements \citep{Sheeley1997_coronal_V, 2002_coronal_electron_density, imamura2014, bemporad2017, casti2023}. To facilitate further use of bipoly modeling by the solar community, a Python version of the code that runs the solutions presented here is available at \hyperref[https://github.com/jbdakeyo/BipolySolarWind]{\url{https://github.com/jbdakeyo/BipolySolarWind}}. 

One might wonder exactly up to what coronal height the bipoly models are constrained by the freeze-in temperatures. Indeed, for a given source region in the low corona, the plasma species are in thermal equilibrium up to the freeze-in height due to the collisional regime (i.e., protons and electrons have the same temperature). Consequently, the determination of the freeze-in height would determine a lower limit height for the coherence of bipoly models. This extension would give the coronal height at which temperatures should be set, which would be an interesting addition to the present study.

\acknowledgments{We acknowledge the NASA Parker Solar Probe Mission and the SWEAP team led by J. Kasper for use of data. We thank the instrumental Solar Wind Analyser team (SWA) for valuable discussions. This research was funded by the European Research Council ERC SLOW\_SOURCE (DLV-819189) project. This work was supported by CNRS Occitanie Ouest and LESIA. We recognise the collaborative and open nature of knowledge creation and dissemination, under the control of the academic community as expressed by Camille Noûs at http://www.cogitamus.fr/indexen.html. 
}

%\bibliography{bibliography}{}
%\bibliographystyle{aasjournal}

\appendix

\section{Isopoly model parameters accounting charge-state ratios observations}
\label{sec:appendix_isopoly_param_table}

\begin{table*}[h]
\hspace{1.5cm}
\scalebox{1}{
%\hspace{-1.7cm}
\begin{tabular}{cccccc}
    %%%%%%%%%%%%%%%%%%%%%%%%%%%%%%%%%%%%%%%%%%%%%%%%%%%
    \hline %\hline 
    Wind type & A & B & C & D & E    
     \tabularnewline
    \hline 
    %%%%%%%%%%%%%%%%%%%%%%%%%%%%%%%%%%%%%%%%%%%%%%%%%%%
    $\Tpo$ (MK) & 1.63 & 1.52 & 1.44 & 1.25 & 1.20
     \tabularnewline
    %\hline 
    %%%%%%%%%%%%%%%%%%%%%%%%%%%%%%%%%%%%%%%%%%%%%%%%%%%
    $\Teo$ (MK) & 1.63 & 1.52 & 1.44 & 1.25 & 1.20
     \tabularnewline
    \hline 
    %%%%%%%%%%%%%%%%%%%%%%%%%%%%%%%%%%%%%%%%%%%%%%%%%%%
    $\risop$ ($\rs$) & 4.5 & 10.5 & 13.6 & 19 & 23
     \tabularnewline
    %\hline 
    %%%%%%%%%%%%%%%%%%%%%%%%%%%%%%%%%%%%%%%%%%%%%%%%%%%
    $\risoe$ ($\rs$) & 2.50 & 2.20 & 3.00  & 2.50 & 2.05
     \tabularnewline
     %%%%%%%%%%%%%%%%%%%%%%%%%%%%%%%%%%%%%%%%%%%%%%%%%%%
    $r_c$ ($\rs$) & 1.38 - 1.29 & 1.34 - 1.29 & 1.74 - 1.26 & 4.62 - 1.26 & 1.26 - 1.29
     \tabularnewline
    \hline 
    %%%%%%%%%%%%%%%%%%%%%%%%%%%%%%%%%%%%%%%%%%%%%%%%%%%
    $\gamp$ & 1.57 & 1.59 & 1.52 & 1.44 & 1.35
     \tabularnewline
    %\hline  
    %%%%%%%%%%%%%%%%%%%%%%%%%%%%%%%%%%%%%%%%%%%%%%%%%%%
    $\game$ & 1.29 & 1.24 & 1.23 & 1.23 & 1.21  
     \tabularnewline
    \hline  
    %%%%%%%%%%%%%%%%%%%%%%%%%%%%%%%%%%%%%%%%%%%%%%%%%%%
    $u_{0}$ (km/s) & x - 45  & x - 39 & x - 37 & x - 27 & x - 23
     \tabularnewline
    %\hline  
    %%%%%%%%%%%%%%%%%%%%%%%%%%%%%%%%%%%%%%%%%%%%%%%%%%%
    $u_{1 au}$ (km/s) & x - 605 & x - 582 & x - 561 & x - 494 & x - 477  
     \tabularnewline
    %%%%%%%%%%%%%%%%%%%%%%%%%%%%%%%%%%%%%%%%%%%%%%%%%%%
    \hline
\end{tabular}  
}
\caption{Updated isopoly input parameters ($\Tpo$, $\Teo$, $\risop$, $\risoe$, $\gamp$, $\game$) associated with the isopoly curves considering expansion factor modeling and $\Teo$ derived from charge-state ratios measured by SolO. The coronal temperature $\Tpo$ and $\Teo$ are shown on the top two lines, the $\risop$, $\risoe$ and critical radius on the three middle lines, the interplanetary polytropic indexes $\gamp$, $\game$, and the derived coronal and 1~au velocities ($u_0$, $u_{1 au}$) on the bottom two lines. The model incorporates the expansion profile $f(r)$ obtained with PFSS and the $\Teo$ derived in Figure~\ref{fig_v_wind_state_charge_equiv_Te}, for each of the 5 wind populations.
The parameters $\risop$ and $\risoe$ have been adjusted to fit the in-situ measured temperatures, velocities. When two values are displayed for a given parameter, the left and right values correspond to 
$\fss<7$ and $100 < \fss < 250$ bins, respectively (where $\fss$ is the flux tube expansion factor at the source surface). 
In the case of a single value, both bins share the same value.
The "x" for a given population means the absence of wind solution. 
}
\label{tab_updated_isopoly_param_modif_Te0}
\end{table*}

\end{document}